\documentclass[runningheads,a4paper]{llncs}

\usepackage{amssymb}
\usepackage{graphicx}
\usepackage{menukeys} 

\usepackage{listings}

\usepackage{color}

\definecolor{dkgreen}{rgb}{0,0.6,0}
\definecolor{gray}{rgb}{0.5,0.5,0.5}
\definecolor{mauve}{rgb}{0.58,0,0.82}

\graphicspath{{./figures/}}

\newcommand{\keywords}[1]{\par\addvspace\baselineskip
\noindent\keywordname\enspace\ignorespaces#1}

\newcommand{\R}{{\mathbb R}}

\begin{document}

\mainmatter

\title{Point Process Models for Distribution of \\ Cell Phone Antennas}

\titlerunning{Point Process Models for Distribution of Cell Phone Antennas}

\author{Ezequiel Fattori\inst{1} \and Pablo Groisman\inst{1} \and Carlos Sarraute\inst{2}}

\institute{Facultad de Ciencias Exactas y Naturales,\\ Universidad de Buenos Aires, Argentina
\and
Grandata Labs, Argentina}

% \authorrunning{AGRANDA 2016, 2º Simposio Argentino de Grandes Datos}
\authorrunning{Fattori, Groisman, Sarraute}

\maketitle

\begin{abstract}
We introduce a model for the spatial distribution of cell phone antennas in 
a urban environment.
After showing that the {\em complete spatial randomness} (homogeneous 
Poisson distribution) hypothesis does not hold, we propose a model in which each 
point is distributed according to a bivariate Gaussian variable with mean given 
by the barycenter of its neighbors in the Delaunay triangulation. 
We show that this model is suitable, and can be used to generate a synthetic distribution of antennas.
The generated distribution contains no sensitive or proprietary information,
and can thus be freely shared with research groups, 
fostering further research on the subject.

\end{abstract}

\section{Introduction}
The scientific analysis of mobile phone datasets, collected by operators within their network, is a recent field of study, with a corpus of published works starting in 2005, and a noticeable increase of publications from 2012 onwards~\cite{naboulsi2015mobile}. Other milestones of this increased interest of the scientific community are the NetMob conferences, focused on mobile phone data analysis~\cite{netmob}.

Mobile traffic contains information about the movement, interactions, and mobile service consumption of individuals at unprecedented scales.
This attracted scientists from multiple disciplines: 
sociologists, epidemiologists, physicists, transportation and telecommunication experts
found in these datasets a clear opportunity to bring their analyses to an unprecedented scale while retaining a high level of detail on each individual.

Thanks to the growing availability of datasets, collaborations between academic research groups and network operators based on the analysis of real-world mobile traffic have been flourishing
(a significant example is the Data for Development (D4D) Challenges by Orange~\cite{d4d}).

One of the main research subject is the mobility analysis. 
Call locations (recorded through cell tower usages) are a projection of users trajectories, 
showing a strong regularity in their movement patterns,
both in space and time~\cite{gonzalez2008understanding}.
Human mobility is also influenced by social components~\cite{cho2011friendship,ponieman2016mobility}.

An important problem of the data collection process is the preservation of subscriber privacy. Mobile traffic data contains sensitive information on individual subscribers,
whose privacy needs to be properly protected~\cite{bigdataandprivacy}.
There are limitations to the current approaches to the anonymization of mobile traffic datasets,
as shown by~\cite{de2013unique}.

One promising approach to avoid privacy issues is to generate synthetic traffic datasets.
Realistic traffic records that do not contain personal information of subscriber can be freely distributed, and can be used by any research entity needing to
perform network simulations and analysis~\cite{Oliveira2015measurement}.

In this work, we tackle the problem of generating a synthetic distribution of antennas, which can be used as a foundation for a synthetic mobile phone traffic generator.
This opens the possibility of sharing (synthetic) datasets,
expanding the access of researchers to the study of mobile phone networks.

Several articles deal with the complete spatial randomness hypothesis 
\cite{Haenggietal1,Haenggi,BaBlas,StoyanKendallMecke} to model the distribution of antennas in cellular phone 
networks, mostly due to its tractability. Although, it has been shown that this kind of point processes is not suitable in realistic 
situations~\cite{GuoHaenggi}. Typical alternatives suggest the use of Mat\'ern processes, determinantal processes, Poisson hard-core processes, Strauss processes, and Perturbed triangular lattice models, but just some of them have been contrasted with real data~\cite{GuoHaenggi}. 

In this article we use real data to test the homogeneous as well as non-homogeneous 
\textit{Poisson point process} (PPP) hypothesis and we conclude that none of them are realistic models for the locations of antennas. As a consequence we propose a new model based on the Harmonic deformation of the Delaunay triangulation of a PPP~\cite{paperpatu} and we contrast the model with real data. This model turns out to be suitable.

\section{Complete Spatial Randomness}
\label{CSR}
The natural mathematical object to deal with in order to model the distribution of antennas in a cellular network is a {\em point process}. A point process is a random variable $X\colon \Omega \to \mathcal N$ that takes values in the space of locally finite configurations (configurations of points in the plane with no accumulation points).

The first thing we would like to do when we face an instance of a point process is to compare it with a benchmark or null model which exhibits \emph{complete spatial randomness} (CSR). By CSR we understand the fulfilment of two conditions:
\begin{itemize}
\item[$\bullet$] Stationarity: The distribution of the process is \emph{translation invariant}, that is, $X$ and $X + p$ have both the same distribution.
\item[$\bullet$] Independent scattering: The number of points in each of $k$ disjoint sets of the domain form $k$ independent random variables.
\end{itemize}

These two conditions guarantee the maximum degree of randomness a process may have, in the sense of the null interaction between points and the homogeneity along the domain, that is, the expected number of points occurring in a subset $S$ is the same as in $S + p$. % \\

It can be proved that if a point process verifies these two conditions then it must be an homogeneous PPP, defined below.

\begin{definition}
Given a point process $X$ and a subset $S \subset \mathbb{R}^2$, $N(S)$ denotes the number of points of $X$ contained in $S$ and $\mu(S):=E(N(S))$ the {\em intensity measure}. We say that $X$ is distributed according to a PPP of intensity $\rho\colon\R~\to~\R_{\ge 0}$ if the following properties are satisfied:
	\begin{enumerate}
	        \item $\mu(S)=\int_S \rho(x)\,dx$ for any square $S=[a,b]\times[c,d]$.
		\item For any square $S$ with $\mu(S) < \infty$, $N(S) \sim \mathcal P(\mu(S))$. The number of points in $S$ is distributed according to a Poisson distribution with mean $\mu(S)$.
		\item For any $n$ and any square $S$ with $0 < \mu(S) < \infty$, conditional on $N(S) = n$, the $n$ points within $S$ are independent and each of them has density $\rho(\cdot) / \mu(S)$.
	\end{enumerate}
When $\rho(x)\equiv \rho >0$ is a constant function we say that $X$ is an {homogeneous Poisson point process}.
\end{definition}

The meaning of the Poisson distribution as distribution of the number of points occurring in $S$ can be realized intuitively as the result of a limit of a binomial distributions. If a grid of thickness $1/n$ is drawn on $\mathbb{R}^2$, and to each square $Q$ of the grid a probability equal to its area is associated for the occurrence of a point within it, then for a subset $S$ consisting of several squares the distribution of the count of points occurring inside $S$ is binomial:
$$	P(N(S) = k) = {{\#Q}\choose{k}} . (n^{-2})^k . (1 - n^{-2})^{\#Q-k}$$
Then, as $n \rightarrow \infty$, and the quotient $\#Q . n^{-2} \rightarrow Area(S)$, by the Poisson limit theorem, this binomial distribution tends to a Poisson distribution with mean $\lambda = \#Q .n^{-2} = Area(S)$.

\begin{figure}
\centering
{\includegraphics[height=4cm]{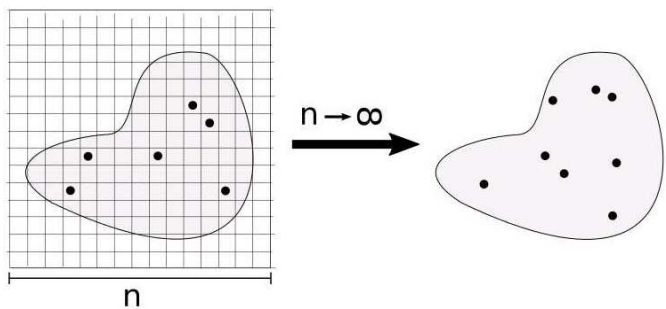}}
  \caption{Poisson point process realized as a limit of binomial random variables.}
\label{fig:grilla_lim}
\end{figure}

\begin{remark}
It must be noted that the CSR hypothesis holds if and only if $\rho$ is a constant since it is the only way in which
	\begin{equation}
		E(N(S + p)) = \mu(S + p) = \mu(S) \quad \mbox{for every } S \mbox{ and } p.		
	\end{equation}
\end{remark}

\section{Nonparametric Estimation of Intensity Functions}

In the case of a homogeneous Poisson process contained in a bounded set $S$, there is a natural unbiased estimate of its intensity, which is in fact the maximum likelihood estimate for a homogeneous Poisson process:
	\begin{equation}
		\hat{\rho} = N(S)/Area(S)	
	\end{equation}

The case of our cellular network is clearly different, since a greater concentration of antennas is observed at least at the city centre. In the cases where homogeneity is suspected not to hold, a non-parametric kernel estimate of the intensity function should be used:
	\begin{equation}
		\hat{\rho}(x) = \sum_{ \zeta \in X} k_b(x - \zeta)	
	\end{equation}

Here $k_b$ represents a volume preserving scaling of a \emph{kernel function} $k$ (which could be a multivariate Gaussian) such that
	\begin{equation}
		\int_{\mathbb{R}^2} k(x) dx = 1	, k(x) \geq 0
	\end{equation}	
	\begin{equation}
		k_b(x) = (1/b)^n k(x/b) .
	\end{equation}
		
This estimate is usually sensitive to the choice of the bandwidth $b$, while the choice of $k$ is less important. We would like to choose the bandwidth $b$ in order to improve the estimation. Namely, the quantity
	\begin{equation}
		\|\rho - \hat{\rho}_b\|_2 = \left(\int_{\mathbb{R}^2} (\rho(x) - \hat{\rho}_b(x))^2 dx \right)^{\frac{1}{2}}
	\end{equation} which measures the distance or dissimilarity between the real intensity $\rho$, which has generated the data, and the estimate $\hat{\rho}_b$ is to be minimized. A good choice for $b$ is the one that minimizes the following quantity:
	\begin{equation}
		L(\hat{\rho}_b) = \|\hat{\rho}_b\|_2^2 - 2 \int_{\mathbb{R}^2} \rho(x) \hat{\rho}_b(x) dx .
	\end{equation}
	
Since this last risk formula involves $\rho$, which is unknown, it can only be estimated. In this case the estimator will be:
	\begin{equation}
		\hat{L}(\hat{\rho}_b) = \frac{1}{|X|} \sum_{\zeta \in X} \left[\|\hat{\rho}_b^{\zeta}\|_2^2 - 2 \hat{\rho}_b^{\zeta}(\zeta) \right]  
	\end{equation} where $\hat{\rho}_b^{\zeta}$ denotes the intensity estimator built from $X - \{\zeta\}$.\\
	
This estimator is known as \emph{Left One Out Cross Validation} (LOO) and it is studied in detail by Celisse \cite{celisse}. Although the author is not able to give a closed formula for the variance and the bias of the estimator, these can be approximately calculated by bootstrapping. This approach, despite being more correct, will not be followed in the present exposition, and instead we will settle for simply choosing the $b$ with less estimated risk.

\begin{figure}[h]
\centering
{\includegraphics[height=7cm]{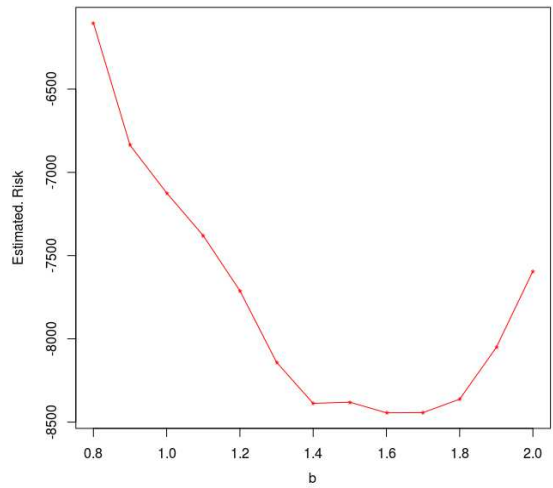}}
\caption{Estimated risk $\hat{L}(\hat{\rho}_b)$ as function of $b$.}
\label{fig:graf_riesgo}
\end{figure}

\begin{figure}[h]
\centering
\makebox[0pt]{\includegraphics[width=\linewidth]{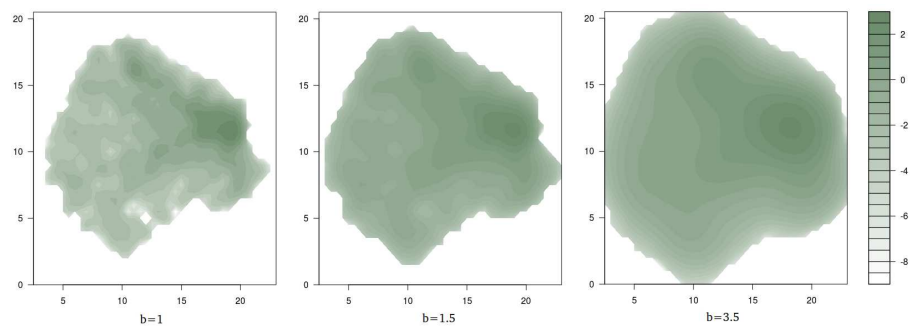}}
\caption{Contour plots of intensity function $\hat{\rho}_b$ in logaritmic scale.}
\label{3intensidades}
\end{figure}

Figure~\ref{fig:graf_riesgo} displays a plot of estimated risk vs. $b$ parameter, showing that the estimated optimal value of $b$ lies around $1.6$ in the chosen scale units. Figure \ref{3intensidades} displays three different bandwidth choices for the estimate of the intensity around the city of Buenos Aires (CABA).

\section{Simulation of Poisson Point Processes}

By simulation of a point process, we understand the generation of an instance of the process drawn from its distribution function. For the same process there may exist various possible procedures or algorithms that simulate it.

The simplest case is the simulation of a homogeneous Poisson process in a box. Let $\rho$ be its intensity and $[a,b]\times[c,d]$ a box in $\mathbb{R}^2$. According to Section \ref{CSR}, the number of point occurrences must be first drawn from a Poisson distribution with mean $\rho . (b-a) . (d-c)$. Once the number $n$ of occurrences is established, the process is simply the result of putting together $n$ independent draws from the uniform distribution $U([a,b]\times[c,d])$. The R code for carrying out these steps is:

%\begin{verbatim}
\begin{lstlisting}[language=R]
   n = rpois(1,lambda=rho*(b-a)*(d-c))
   pois_sim = matrix(0,2,n)
   for(i in seq(1,n)){
      pois_sim[,i] = c(runif(1,a,b),runif(1,c,d))
   }
\end{lstlisting}
%\end{verbatim}

The simulation of a non-homogeneous process is less simple. In order to do this, the box is first partitioned into a grid of $m \times m$ rectangles, in each of whom the intensity function is assumed to be approximately constant. Since these rectangles are pairwise disjoint, they form $m \times m$ independent Poisson processes, each of whom is simulated like the simple case handled above. The R code for carrying on these steps is:

%\begin{verbatim}
\begin{lstlisting}[language=R]
   pois_sim = matrix(0,2,0)
   for(i in seq(1,m)){
      for(j in seq(1,m)){
         n = rpois(1,lambda=rho((b-a)/m*(j-1), ~
            (d-c)/m*(i-1))*(b-a)/m*(d-c)/m)
         pois_sim_sub = matrix(0,2,n)		
         for(k in seq(1,n)){
            pois_sim_sub[,k] = c(runif(1,(b-a)/m*(j-1),(b-a)/m*j), ~
               runif(1,(d-c)/m*(i-1),(d-c)/m*i))
         }
         pois_sim = cbind(pois_sim,pois_sim_sub)
      }
   }
\end{lstlisting}

Last but not least, there exist a R library written by Adrian Baddeley called \emph{spatstat} intended to handle spatial point processes. This library contains a function which simulates non-homogeneous Poisson processes taking by argument its intensity function and returning an object of type ``ppp". An R code example of its use for generating a Poisson process with intensity $\rho$ in $[0,1]\times[0,1]$ is displayed below:

%\begin{verbatim}
\begin{lstlisting}[language=R]
   window = owin(c(0,1),c(0,1))
   lmax = 100
   pois_sim_ppp = rpoispp(rho,lmax,window)
   pois_sim = rbind(pois_sim_ppp$x,pois_sim_ppp$y)
\end{lstlisting}

\section{Complete Spatial Randomness Testing}

Testing the CSR hypothesis is an important part of the analysis of point patterns. If the hypothesis is accepted, then it is not possible to find indicators of interesting interaction between the points based on the geometry observed in the pattern. This analysis provides also information on the direction of the deviation from CSR, as well as its cause to suggest further analysis.% \\

In our present case, we suspect at first hand that the process of antennas is not CSR, because it has a clear non-homogeneity. Nevertheless, it is still necessary to determine if the non-homogeneous Poisson hypothesis holds. In order to do this, we proceed as follows. Under the Poisson hypothesis, if we apply an adequate \emph{thinning transformation} to the process, the resulting process should be Poisson homogeneous, that is, CSR. Then we can apply a test for CSR, and if the test rejects CSR for the transformed data, then it also rejects the Poisson hypothesis for the original data.

\begin{figure}
\centering
{\includegraphics[width=\linewidth,height=5cm,keepaspectratio]{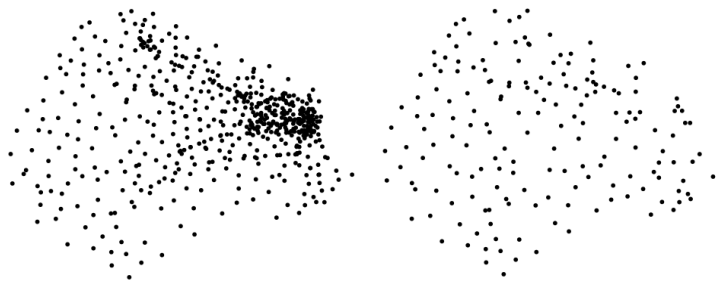}}
\caption{Distribution of cellphone antennas before and after homogenization.}
\label{caba_antes_desp}
\end{figure}

A thinning transformation consists of a random drop of points of the process. In the case of location-dependent thinning, a function $p(x)$ with range $[0,1]$ is defined, and each point $x$ is dropped with probability $1 - p(x)$. The result of applying a thinning operation to a Poisson process is also a Poisson process. If the intensity of the original process is $\lambda(x)$, the new intensity becomes $\lambda(x) . p(x)$. So that if we apply $p(x) = 1/\lambda(x)$, assuming $\lambda(x)>1$ everywhere, the process becomes homogeneous. \\

Many statistics for testing CSR are available, one of the most important is \emph{Ripley's K function}. For stationary isotropic processes the $K$ function is defined as:
	\begin{equation}
		K(t) = \lambda^{-1} E[N_0(t)]
	\end{equation}where $N_0(t)$ represents the total count of points of $X$ at distance less than $t$ from the origin, which, disregarding edge effects, in the Poisson homogeneous case should be the same as $N_p(t)$ for any $p$. 

A first estimate that can be devised for this quantity is the following:
	\begin{equation}
		\hat{K}(t) = \frac{1}{\lambda n}  \sum_i \sum_{j \neq i} I(r_{ij} < t)
	\end{equation} where $I$ is the indicator function and $r_{ij}$ is the distance between $x_i$ and $x_j$. However, this estimate is negatively biased because of edge-effects. For a point within distance $t$ of the boundary of $S$, the observed count of other points within distance $t$ necessarily excludes any events which may have occurred within distance $t$ but outside $S$. Several estimates have been proposed to correct this bias, the following is due to Ripley:
	\begin{equation}
		\hat{K}(t) = \frac{1}{\lambda n}  \sum_i \sum_{j \neq i} I(r_{ij} < t) w_{ij}^{-1}
	\end{equation} where $w(x,r)$ stands for the proportion of the disk $D(x,r)$ contained in $S$, and $w_{ij}$ represents $w(x_i,r_{ij})$.  \\

\begin{figure}[h]
\centering
\includegraphics[width=0.7\linewidth,height=7cm,keepaspectratio]{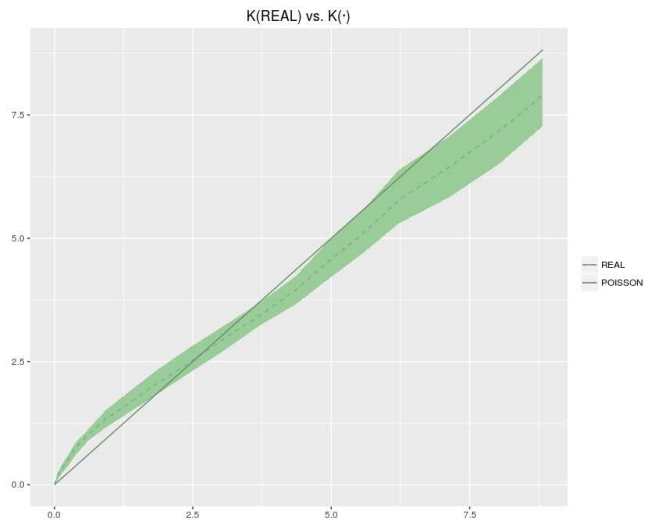}
\caption{Empirical Ripley's $K$ function of the actual distribution, theoretical function for Poisson homogeneous processes (dashed) and min-max envelopes for 100 simulated Poisson homogeneous processes (shaded).}
\label{fig:Kpoisson}
\end{figure}

Using the last estimate, the $K$ function of our \emph{thinned} process is compared with the (theoretically calculated) expectation of a Poisson homogeneous process. The results are displayed 
in Fig.~\ref{fig:Kpoisson} in the form of a parametric $K$ vs. $K$ plot, where $t$ is the parameter, the value of $K(t)$ for the thinned process is the $x$ coordinate and the value of $K(t)$ for any other process (indicated in the legend) is the $y$ coordinate. This modality of plot is proposed by Diggle \cite{diggle}.

In the plot the $K$ function of the actual thinned process is the black identity curve, and the dashed line is the theoretically predicted function, according to the formula $K(t) = \pi t^2$.
%\begin{equation}
%	K(t) = \pi t^2
%\end{equation}

The shaded area under the dashed line represents a \emph{min-max envelope}. An envelope is a curve which is constructed from a set of curves by means of some type of operation. The minimum and maximum envelopes are constructed from the empirical $K$ functions of simulated Poisson processes according to:
\begin{equation}
	U(t) = max\{K_1(t), K_2(t), \dots\} , L(t) = min\{K_1(t), K_2(t), \dots\} 
\end{equation} and the shaded area is the region between both curves. The goal of constructing an envelope is to represent the dispersion of the empirical functions around their calculated expected value for each $t$.

In our case, the envelope is bent upward for small values of $t$, that is, the process lacks small distances between points, which indicates a departure from the Poisson hypothesis. Our process exhibits \emph{repulsion} between points.

\section{Harmonic Deformation of Delaunay Triangulation}

In a recently published paper, Ferrari, Groisman and Grisi \cite{paperpatu} proved the existence and constructed a point process model that seems to fit well the distribution of antennas.

Let $X$ be an homogeneous PPP in $\R^2$. The \emph{Voronoi cell} associated to each point $p$ of $X$ is the region made up of all points in $\R^2$ which are closer to $p$ than to any other point of $X$. The \emph{Voronoi neighbors} of $p$ are all the points of $X$ other than $p$ whose Voronoi cell's closure intersects the closure of the one of $p$, that is, the owners of the adjacent cells. The \emph{Delaunay triangulation} of $X$ is the graph built from all points of $X$ and all edges made up of pairs of Voronoi neighbors.

Let $H: X \rightarrow \R^2$ be a function with the properties that for each $p$ in $X$, $H(p)$ is located at the barycenter of  $\{ H(q) | q \leftrightarrow p \}$.
Here $q \leftrightarrow p$ denotes that $p$ and $q$ are Delaunay neighbors.

\begin{figure}[t]
\centering
{\includegraphics[width=1.0\linewidth,height=5cm,keepaspectratio,trim={1cm 0.5cm 1cm 0.5cm},clip=true]{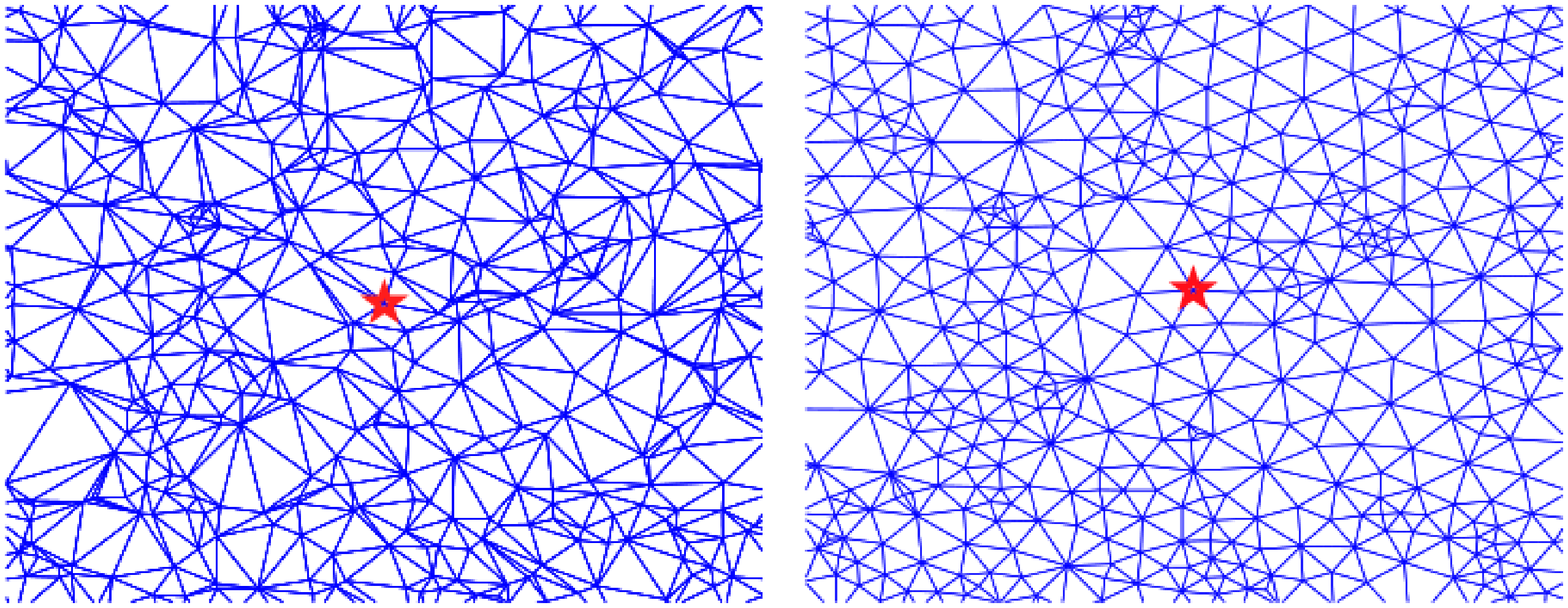}}
\caption{Delaunay triangulation of an homogeneous PPP (left) and its harmonic deformation (right). The star indicates the origin (left) and $H(0)$ (right).}
\label{harmonic_def}
\end{figure}

Such a function is called a ``Harmonic function", and its existence was proved in the above cited paper. It represents a position change made to every point of $X$ such that every point is relocated in the barycenter of its, also relocated, Delaunay neighbors (as illustrated by Fig.~\ref{harmonic_def}).\footnote{The two illustrations of Fig.~\ref{harmonic_def} were extracted from Ferrari, Groisman and Grisi \cite{paperpatu}.}

This original model must be in our case somehow adapted to fulfil our needs. The main idea is to simulate a non-homogeneous PPP according to our estimated intensity $\hat{\rho}$ for the city of Buenos Aires (CABA), calculate its Delaunay triangulation and finally apply a function similar to $H$ to locate every point at the barycenter of its neighbors. Since our region is bounded, we cannot let the condition (1) hold for every point because the only possible solution would be that all points were collapsed to an unique point. It is therefore necessary to allow some points to not meet this condition, and to keep them fixed by $H$ instead.\\

Which points should be kept fixed? Many recipes were tested empirically: letting too few fixed points leads to an unwanted deformation of the intensity of the resulting process; letting too many fixed points fails to deviate from the Poisson model. Ultimately, we would like to perform the harmonic deformation over our bounded region $S$, without fixing any point and without altering the original intensity function. Our proposed solution is therefore to:
\begin{itemize}
	\item Extend the intensity function smoothly over a box $B$ containing region $S$.
	\item Simulate a PPP according to this intensity in $B\setminus S$.
	\item Join these points with the ones in $S$.
	\item Perform an harmonic deformation in $B$ leaving the points in $B\setminus S$ fixed.
\end{itemize}

In order to smoothly extend, in an approximate manner, the intensity function, there is a simple approach that consists in dividing the box $B$ into a grid and assign to each intersection of grid lines a variable $\rho_{ij}$. For those intersections falling inside $S$, the variables are fixed to the actual value of the intensity function. For those intersections falling in $B\setminus S$ the values are to be determined; and for those falling in the border of $B$ the values are fixed to zero. The linear equations:
\begin{equation}
	\rho_{ij} = ( \rho_{i+1,j} + \rho_{i,j+1} + \rho_{i-1,j} + \rho_{i,j-1} ) / 4
\end{equation} hold for each intersection in $B\setminus S$, the resulting system is determined and can be solved by matrix inversion or any other more efficient algorithm. Once all the values of the extended function at the intersections are found, the value of the function in an arbitrary point of the box $B$ can be set to match those of the nearest grid intersection. This corresponds to the harmonic extension of $\rho$ from $S$ to $B$.

\begin{figure}
\centering
{\includegraphics[width=0.95\linewidth,height=7cm,keepaspectratio]{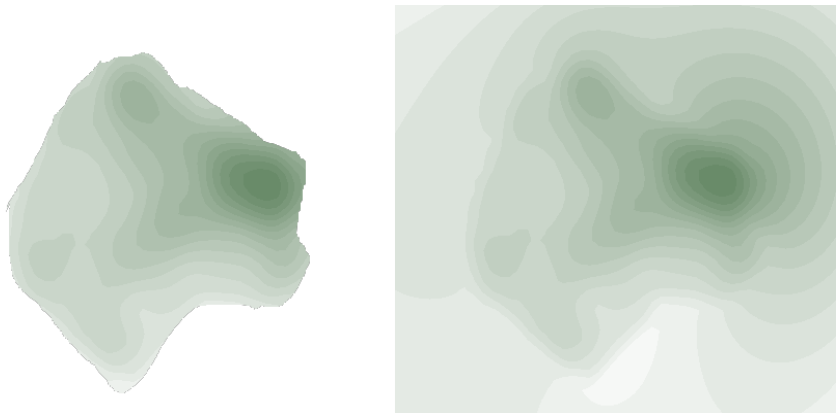}}
\caption{Smooth extension of the intensity function to a box containing the city of Buenos Aires.}
\label{intensity_ext}
\end{figure}

The harmonic deformation can be achieved by an iterative procedure. First the Delaunay triangulation is calculated for the whole box $B$, consisting of the original points belonging to $S$ and the simulated ones. Any point may have several neighbors from the two classes. A Delaunay triangulation function ``delaunayn()" is built in the library ``geometry" from R.
Secondly the points within $S$ are copied into a sequence. Then an iteration is made over this sequence, where each step consists of relocating the present element at the barycenter of its neighbors. Thus, the points in $B\setminus S$ are always fixed and the ones in $S$ change within each iteration and are stored in the sequence. This iteration is performed several times until some prescribed degree of convergence is achieved. \\

This sole model is better than the Poisson model but its is not enough by its own, because it is not realistic. In the actual distribution, each antenna is not located exactly at the barycenter of its neighbors. In order to make the model fit better the real distribution of antennas, we will add a \emph{noise parameter $\epsilon$} so that for every $x \in X$ the following condition hold:

	\begin{equation}
		x|\{y \in X | y \leftrightarrow x\} \sim N_2(\mu,\epsilon(\mu)I)
	\end{equation}

	\begin{equation}
		\mu = \frac{1}{n_x} \sum_{y \leftrightarrow x}y
	\end{equation} where $n_x$ is the number of neighbors of $x$. This means that for each $x \in X$, conditioned on the position of its neighbors, the position of $x$ is Gaussian with mean the barycenter $\mu$ of the neighbors and variance equal to $\epsilon$. % \\ \\ \
	A more detailed study of $\epsilon$ as a \emph{function} $\epsilon(x)$ is left to further work. %\\
	
	The following R code implements the complete procedure. In this code we take $\epsilon(x)$ as the square root of the (estimated density) at $x$.

\medskip
% \noindent

% \begin{verbatim}
\begin{lstlisting}[language=R]
#build a neighborhood matrix from the output of "delaunayn()"
#applied to the simulated poisson process "pois_sim" in B.

#"pois_sim" is built from the simulated points in S and c(S)
count_S <- ncol(pois_sim_S)
pois_sim <- cbind(pois_sim_S,pois_sim_cS)
count_B <- ncol(pois_sim)

neighb_matrix <- matrix(0,count_B,count_B)
simplexes <- delaunayn(t(pois_sim))
for (i in seq(1,nrow(simplexes))){
   neighb_matrix[simplexes[i,1],simplexes[i,2]] <- 1
   neighb_matrix[simplexes[i,2],simplexes[i,1]] <- 1
   neighb_matrix[simplexes[i,1],simplexes[i,3]] <- 1
   neighb_matrix[simplexes[i,3],simplexes[i,1]] <- 1
   neighb_matrix[simplexes[i,3],simplexes[i,2]] <- 1
   neighb_matrix[simplexes[i,2],simplexes[i,3]] <- 1
} 

#iterate the relocation procedure until each point x moves
#less than k.epsilon(x). In this example k=3.
done <- FALSE
while(done == FALSE){
   done <- TRUE
   for(i in seq(1,count_S)){
      relocated_point <- c(0,0)
      count_neighb <- 0
      for(j in seq(1,count_B)){
         relocated_point <- relocated_point + ~
                        pois_sim[,j] * neighb_matrix[i,j]
         count_neighb <- count_neighb + 1 * neighb_matrix[i,j]
      }
      relocated_point <- relocated_point / count_neighb
      #once the point is in the barycenter of neighbors, add noise
      relocated_point <- relocated_point + ~
                        rnorm(n=2,mean=0,sd=epsilon(relocated_point))
      if(sqrt(sum((relocated_point - pois_sim[,i])^2)) > ~
                        3*epsilon(pois_sim[,i])){
         done <- FALSE
      }
      pois_sim[,i] <- relocated_point
   }
}
\end{lstlisting}

\begin{figure}
\centering
{\includegraphics[width=\linewidth]{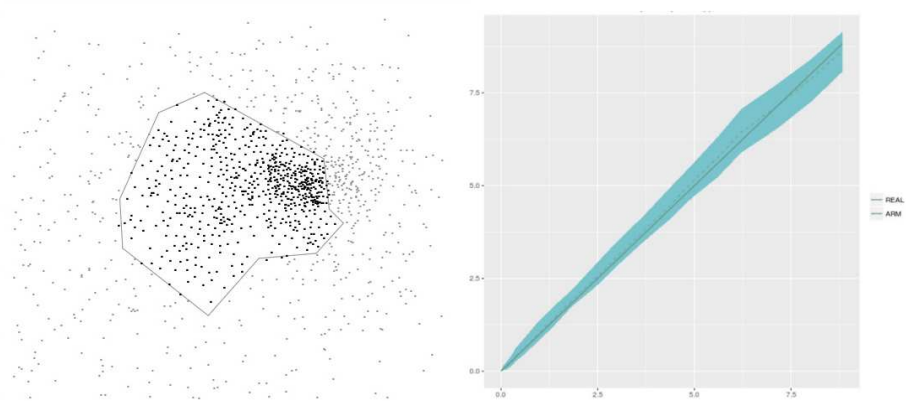}}
\caption{Example of the final model (left). Upper and lower envelopes of Ripley’s $K$ function for 100 simulated (and homogenized) processes from the harmonic deformation with noise model (right).}
\label{resultado_harm}
\end{figure}

Figure~\ref{resultado_harm} displays the final result and assessment of the harmonic deformation with noise model via the Ripley's K function. The left image also includes the auxiliary points from the simulated PPP in $B\setminus S$ (grey) used to construct the model.
As it can be seen on the image, there is no significant deviation between the $K$ function of the actual distribution of antennas and the final simulated model.

\section{Conclusions}

Using real cell tower location data, we showed that a Poisson point process is not a realistic model for the distribution of antennas in cellular phone networks. 
We proposed an alternative point process in which each point is (conditioned on the positions of its neighbors) distributed according to a bivariate Gaussian distribution centered at the barycenter of the neighbors. 
Several models can be used to describe the variance of these Gaussian distributions. We are still investigating which is the most suitable.

We used real data for the location of antennas in a urban environment to show that this model is adequate for this situation. The extension to other cities or regions is planed as future work.
This model can be used to generate a realistic spatial distribution of antennas,
or to simulate the growth of the network (i.e. the increase in number of antennas).
Since the real location of antennas is sensitive and proprietary information,
the possibility of generating a synthetic distribution is also useful in practice:
such distribution can be freely shared with research groups, 
thus fostering collaborations between industry and academic research.

\bibliographystyle{unsrt}
\bibliography{antenas}

\end{document}